# JGR Atmospheres

**RESEARCH ARTICLE**
10.1029/2019JD030312

**Key Points:**
- Gamma ray glows were observed for the first time at 20-km altitude above thunderclouds
- The thunderclouds below the aircraft had an anomalous charge structure
- Possible production mechanisms for gamma ray glow are tested by Monte Carlo modeling

**Supporting Information:**
- Supporting Information S1
- Data Set S1

**Correspondence to:**
N. Ostgaard,
nikolai.ostgaard@uib.no

**Citation:**
Østgaard, N., Christian, H. J., Grove, J. E., Sarria, D., Mezentsev, A., Kochkin, P., et al. (2019). Gamma ray glow observations at 20-km altitude. *Journal of Geophysical Research: Atmospheres*, *124*, 7236–7254. https://doi.org/10.1029/2019JD030312

Received 14 JAN 2019
Accepted 24 APR 2019
Accepted article online 16 MAY 2019
Published online 8 JUL 2019

**Author Contributions:**
**Data curation:** N. Østgaard, H. J. Christian, J. E. Grove, E. Wulf, G. Genov, K. Ullaland, M. Marisaldi, S. Yang, R. J. Blakeslee
**Funding Acquisition:** N. Østgaard, H. J. Christian
**Methodology:** N. Østgaard, D. Sarria, N. Lehtinen
**Software:** N. Østgaard, J. E. Grove, D. Sarria, A. Mezentsev, P. Kochkin, M. Quick, K. Ullaland
**Validation:** N. Østgaard, J. E. Grove, D. Sarria, A. Mezentsev, P. Kochkin, N. Lehtinen, M. Quick, E. Wulf, G. Genov, K. Ullaland, M. Marisaldi, S. Yang, R. J. Blakeslee
**Writing - Original Draft:** N. Østgaard, D. Sarria

# Gamma Ray Glow Observations at 20-km Altitude

N. Østgaard[1], H. J. Christian[2], J. E. Grove[3], D. Sarria[1], A. Mezentsev[1], P. Kochkin[1], N. Lehtinen[1], M. Quick[4], S. Al-Nussirat[5], E. Wulf[3], G. Genov[1], K. Ullaland[1], M. Marisaldi[1], S. Yang[1], and R. J. Blakeslee[4]

[1]Birkeland Centre for Space Science, University of Bergen, Bergen, Norway, [2]Department of Atmospheric Science, University of Alabama, Huntsville, AL, USA, [3]U.S. Naval Research Laboratory, Washington, DC, USA, [4]NASA Marshal Space Flight Center, Huntsville, AL, USA, [5]Department of Physics and Astronomy, Louisiana State University, Baton Rouge, LA, USA

**Abstract** In the spring of 2017 an ER-2 aircraft campaign was undertaken over continental United States to observe energetic radiation from thunderstorms and lightning. The payload consisted of a suite of instruments designed to detect optical signals, electric fields, and gamma rays from lightning. Starting from Georgia, USA, 16 flights were performed, for a total of about 70 flight hours at a cruise altitude of 20 km. Of these, 45 flight hours were over thunderstorm regions. An analysis of two gamma ray glow events that were observed over Colorado at 21:47 UT on 8 May 2017 is presented. We explore the charge structure of the cloud system, as well as possible mechanisms that can produce the gamma ray glows. The thundercloud system we passed during the gamma ray glow observation had strong convection in the core of the cloud system. Electric field measurements combined with radar and radio measurements suggest an inverted charge structure, with an upper negative charge layer and a lower positive charge layer. Based on modeling results, we were not able to unambiguously determine the production mechanism. Possible mechanisms are either an enhancement of cosmic background locally (above or below 20 km) by an electric field below the local threshold or an enhancement of the cosmic background inside the cloud but then with normal polarity and an electric field well above the Relativistic Runaway Electron Avalanche threshold.

## 1. Introduction

Thunderstorms have been shown to produce a variety of hard radiation ranging from the submillisecond long terrestrial gamma ray flashes (Fishman et al., 1994) to the tens of milliseconds long relativistic electron beams (Dwyer et al., 2008) and to the tens of seconds of gamma ray glows.

The first successful experiment to observe enhanced X-ray production during thunderstorms was reported by Parks et al. (1981). At a flight altitude of 4 km they observed a factor ~3 increase of X-rays (>12 keV) above background for 18–20 s. They suggested that their observations could be due to atmospheric bremsstrahlung X-rays from electrons that were produced by runaway processes, as suggested by Wilson (1925). Some years later McCarthy and Parks (1985) reported enhanced fluxes of X-rays (≥5 to ≤110 keV) at 9-km altitude 10–100 times the background flux lasting for about 10 s. This was also the first paper to report that the glow decreased down to background level when a lightning stroke was observed. No electric field measurements were available during these flights. In a modeling paper McCarthy and Parks (1992) revisited these results and tested various models that could explain their observations. As seed particles they considered both cosmic background and radionuclides. They considered electric field strength both below and above the break-even field for relativistic particles, which is now usually termed the Relativistic Runaway Electron Avalanche (RREA) threshold field ($E_t$). They emphasized that their model was simple but nevertheless provided indications that cosmic background is the most probable seed population and that an electric field strength of $>2/3 E_t$ could explain the observed enhancements.

Several balloon experiments were performed over Oklahoma during the 1990s. Eack et al. (1996a) reported observations at 15-km altitude (well above the cloud top at 12 km) of 1-s-long pulses of X-rays (30–120 keV) 10–100 times the background level. No significant electric field increase was seen by their electric field meter. With a similar X-ray detector (30–120 keV). Eack et al. (1996b) reported enhanced X-ray fluxes 2 orders of magnitude above the background level at 4 km lasting for about 1 min. Electric field increase to about 2/3 $E_t$

**Formal Analysis:** N. Østgaard, D. Sarria, P. Kochkin, N. Lehtinen
**Investigation:** N. Østgaard, H. J. Christian, J. E. Grove, D. Sarria, A. Mezentsev, P. Kochkin, M. Quick, E. Wulf, G. Genov, M. Marisaldi, S. Yang, R. J. Blakeslee
**Project Administration:** N. Østgaard, M. Quick, G. Genov, K. Ullaland, R. J. Blakeslee
**Resources:** N. Østgaard, H. J. Christian, G. Genov, K. Ullaland, S. Yang, R. J. Blakeslee
**Supervision:** N. Østgaard
**Visualization:** A. Mezentsev
**Writing - review & editing:** A. Mezentsev, P. Kochkin, N. Lehtinen, G. Genov, M. Marisaldi

coincided with the X-ray enhancements. Eack et al. (2000) reported X-rays increase 3 times the background level for about 20 s at 14 km when the balloon descended through a thunderstorm anvil. These results were revisited by Eack and Beasley (2015) who concluded that the long-duration X-ray emissions they observed were most likely related to production of runaway electrons. However, when available, their electric field measurements never exceeded the $E_t$ but could reach a significant fraction at least 2/3 of the $E_t$.

More recently, gamma ray glows have been observed during a few aircraft campaigns. Observations of gamma ray glows (50 keV to 5 MeV) by the Airborne Detector for Energetic Lightning Emissions were reported by Kelley et al. (2015). During 37 flight hours at a cruising altitude between 14 and 15 km, they detected 12 gamma ray glows. Their durations were from 4 to 112 s, and the brightest glow was up to 1–2 orders of magnitude above the background level (assuming that background is about ∼1,000 counts per second as we inferred from the background variation in their Figure 1) and was abruptly terminated. The authors suggested that the glow was terminated by a (nondetected) lightning stroke. Although no electric field measurements were available, they inferred from modeling of this brightest glow that the field had to be ∼1.5 $E_t$ above the aircraft and concluded that a downward propagating RREA with more than eight avalanche lengths (∼1,000 m) between the negative screening layer and the top of the main positive charge region at 14 km (the flight altitude) was consistent with the measurements. Kochkin et al. (2017) reported two gamma ray glows (100 keV to 10 MeV) observed by the In-Flight Lightning Damage Assessment System when flying inside the thundercloud at 12-km altitude. Both glows lasted ∼30 s, and the gamma ray fluxes reached 20 and 3 times above the background level, respectively. The first glow was abruptly terminated when a distant lightning was detected by the In-Flight Lightning Damage Assessment System. Both Dwyer et al. (2015) and Kochkin et al. (2018) have reported short pulses (200 ms and −1 s) of enhanced fluxes of 511-keV emissions, indicating an enhanced flux of positrons annihilating. How the enhanced fluxes of positrons are produced is still a mystery. However, in both cases these short pulses were seen on top of longer-lasting (1–2 min) weak gamma ray glows (2–3 times the background) and both observations were obtained inside an active thundercloud, ∼14 and 12 km, respectively. The short pulses were observed simultaneously with the dumping of small controlled portions of charge from the aircraft by the static dischargers onboard.

Gamma ray glows, although much weaker, have been frequently observed both on mountaintops and at sea level. Gamma ray enhancements of 50–100% above background level for >1 min were observed at sea level in Japan during winter storms (Tsuchiya et al., 2011, 2013), while Chilingarian et al. (2011) and Chilingarian (2014) reported enhancements usually of a few percentage, but occasionally up to 100%, above the background level that were detected at the Aragats Mountain in Armenia, 3,200 m above sea level, and lasting for several minutes. Chilingarian et al. (2011) termed their observations Thunderstorm Ground Enhancements, and in a few modeling papers (Chilingarian et al., 2012, 2014) they found strong support for the Modification Of Spectra (MOS) process as the production mechanism. This mechanism requires field strength of just a fraction of $E_t$ but gives the cosmic background electron and positron sufficient extra energy to produce enhanced fluxes of gamma rays. The spectral shape resulting from RREA and MOS was also investigated by Cramer et al. (2017) who found similar spectral behavior as reported by Chilingarian et al. (2014).

We should also mention that short pulses (∼1 ms) of X-rays have been observed from the ground (Moore et al., 2001; Dwyer et al., 2004, 2005). These are transients related to leader steps and may have a different origin than the longer-lasting glows.

To summarize this brief review of earlier results, large enhancements (2–3 orders of magnitude) lasting for tens of seconds have been observed inside the thundercloud. Above the thundercloud top the enhanced fluxes that last for tens of seconds are much smaller (factor of ∼3). However, short pulses (∼1 s) of large enhancements (10–100 times the background) have been observed above the cloud top (15 km). On the ground, the enhanced fluxes are even smaller, typically from a few to tens of percents. Only in a few cases have electric field measurements been available and indicate field strength of a significant fraction of the local $E_t$ but not above the $E_t$. Field strengths above $E_t$ have only been inferred from modeling. The long duration of the glows (tens of seconds) indicates a spatial structure, typically 10–20 km, of enhanced electric field strength.

In this paper we report observations of gamma ray glow from an even higher altitude, at 20 km. We will give an overview of the gamma glow observations as well as other data that can help us to understand the thundercloud system and its charge structure over which we were flying. We will also explore different production mechanisms for producing the gamma glow, whether the glow is produced locally (above/below 20 km) or inside the thundercloud (below the cloud top at 13.5 km).

During the glow, there was an abrupt (∼200 ms) decrease in glow intensity. A detailed analysis of this gamma glow reduction will be presented in a separate paper.

## 2. Instruments and Data

In the spring of 2017 the "Geostationary Operational Environmental Satellites (GOES)-R Validation Flight Campaign" was undertaken with an ER-2 aircraft over the continental United States. The scientific targets of the campaign was twofold: (1) to validate the observations by the Advanced Baseline Imager and Geostationary Lightning Mapper (GLM) instruments onboard the recently launched GOES-R satellite: Fly's Eye GLM Simulator (FEGS) and (2) to observe energetic radiation from thunderstorms and lightning: Airborne Lightning Observatory for FEGS and TGFs (ALOFT). The ER-2 aircraft was equipped with a suit of instruments designed for detecting hard radiation from thunderstorms and optical and electric signatures from lightning and multiwavelength measurements of cloud structures. Here we will describe the instruments and the data that are used in this study.

There were two detector systems for detecting X-ray and gamma ray. The University of Bergen (UIB)-Bismuth-Germanium-Oxide (BGO) detector consisted of three BGO scintillator bars (15 cm × 5 cm × 3.2 cm) each one connected to photomultiplier tubes (PMT) and a small LYSO detector (1 cm$^3$) designed to detect high fluxes of gamma rays. As only the UIB-BGO (and in Situ Thunderstorm Observer for Radiation Mechanism [iSTORM]; see description below) detected the gamma ray glow reported here, we will not show any data from the LYSO detector. The measurements from the UIB-BGO detector, which will be shown here, have a 27-ns time resolution and cover the energy range from 300 keV to ∼40 MeV. The three BGO bars were energy calibrated independently on ground using known gamma ray sources and the muon peak, which was modeled by Geant4 simulation to be at 31.7 MeV for a 3.2-cm-thick BGO bar. The UIB-BGO detector and front-end electronics are similar to the high-energy detector on Atmosphere-Space Interactions Monitor (Østgaard et al., 2019), but contrary to Atmosphere-Space Interactions Monitor, which has a triggered system, a data acquisition and storage system was developed for UIB-BGO to enable continuous data recording during the flights.

The other gamma detector system, the iSTORM gamma ray transient detector contains two 2×2×2-in. plastic scintillator cubes each read out by a PMT, one 4-in.-long × 2-in.-diameter CLLB (Cs2LiLaBr6) inorganic scintillator crystal read out by a PMT, and one 1 × 1 × 1-in. plastic scintillator cube read out by a silicon photomultiplier. The three PMTs are read out by a multichannel analyzer digitizing at 120 MHz, while the silicon photomultiplier is read out by an multichannel analyzer at 80 MHz. The signals from all four detectors are time aligned to better than a microsecond. The data shown in this paper are from the two large plastic detectors, which has an energy range of ∼100 keV to ∼8 MeV.

There were three systems for detecting electric fields. In this paper we will only include electric field data from the ER-2 Lightning Instrument Package (LIP). The ER-2 LIP configuration for this flight campaign consists of seven electric field mills (Bateman et al., 2007) and a dual-tube Gerdian conductivity probe. We installed upward and downward facing electric field mills along the centerline of the aircraft aft of the cockpit, as well as along the centerline of each of the ER-2 superpods (two mills on one pod and three on the other). Then, using an iterative matrix analysis (Mach, 2015; Mach & Koshak, 2007), we obtain the vector components of the electric field (i.e., $E_X$, $E_Y$ and $E_Z$) over a wide dynamic range extending from fair weather electric fields (on the order of 10 V/m) to large thunderstorm fields (tens of tens of kilovolts per meter) with a time resolution of 0.1 s and an uncertainty around ±20%. The abrupt electric field changes in the data indicate lightning discharges. Often, it is possible to differentiate between intracloud (IC) and cloud-to-ground (CG) discharges. The vector components of the electric field greatly improve the knowledge and interpretation of the electrical structure of storms overflown, particularly when the ER-2 passes over storms off center or encounters complex or anomalous charge structures.

The Cloud Physics Lidar (CPL) is a backscatter lidar designed to operate simultaneously at three wavelengths: 1,064, 532, and 355 nm. The CPL provides multiwavelength measurements of cirrus, subvisual cirrus, and aerosols with high temporal and spatial resolution. The vertical resolution of the CPL measure-ments is fixed at 30 m; horizontal resolution can vary but is typically about 200 m. In this study the CPL is used to identify the cloud top. Temporal resolution is 5 s.

Data from several lightning detection networks were available during the flight. The World Wide Lightning Location Network (WWLLN) provides lightning geolocation and timing by the use more than 70 VLF sensors around the globe (for more information, see http://wwlln.net). The time of group arrival technique provides average accuracy of 5 km and 10 μs but can vary significantly with geographical origin of the storm (Hutchins et al., 2012; Rodger et al., 2005). To geolocate a lightning event, the WWLLN needs to detect the VLF sferics at least by five stations (Rodger et al., 2005).

The Colorado Lightning Mapping Array (COLMA; Rison et al., 2012) was operating 15 stations during our aircraft observations. The distance between the array center and the glow observations was ∼140 km. From COLMA information about sferics latitude, longitude, altitude, reduced chi-square ($\chi^2_{red}$), and power was obtained. In order to filter out the most reliable sferics, only those with $\chi^2_{red}$ lower than 0.5 were used.

The National Lightning Detection Network (NLDN) provides geolocation of lightning radio pulses and differentiates between IC and CG lightning as well as polarity and peak current. NLDN has good location accuracy and detection efficiency over the United States.

## 3. Observations

Figure 1a shows four convective cloud systems imaged by the GOES 15 Imager satellite (630 nm) in Colorado with lightning activity detected by WWLLN (red dots). ER-2 was flying at 20-km altitude over the two northernmost systems coming in from the east passing over the first system, where two sequences of gamma glows were observed (marked with light gray segments) and then made a right turn to fly northward over the northwest system, where no glow was observed. The inset (Figure 1b) shows the gamma ray observations (counts per second) by the UIB-BGO detector (>300 keV) along the track. Since the aircraft was flying westward, time is running to the left. Figure 1c shows the flight altitude and cloud tops, while Figure 1d shows the observed gamma rays (counts per second) by UIB-BGO during the entire flight that day.

### 3.1. Observations During the Glow

In Figure 2 we zoom in on the observations during the glow events. Figures 2a–2c show the electric field components measured by LIP at 20-km altitude. The coordinate system for the electric field is as follows: $x$ is toward the nose of the aircraft, $z$ is upward, and $y$ to the left of the aircraft. Figure 2d shows the altitude of the cloud top (CPL) while Figure 2e and 2f show the gamma ray observations by UIB-BGO and iSTORM, respectively.

From the UIB-BGO measurements (Figure 2e) we see that the background level is about 2,500 counts per second and during the first passage of the glow the count rate first increased to ∼2,650 counts per second (6% increase) and then for a short interval (15 s) increased to ∼3,650 counts per second (45% increase above background) and then abruptly reduced to 2,900 counts per second (16% above the background). This abrupt (∼200 ms) decrease will be analyzed in detail in a separate paper. At ∼21:48 UT the count rate returned to the background level, but at 21:50 UT a new increase to about 2,750 counts per second (10% increase) was observed. From the time profile one could argue that there might be more than one source for these gamma ray enhancements. The rapid increase and drop in intensity between 21:47:15 and 21:47:30 UT (45% increase in only 15 s) could be from a more localized (or transient) source than the longer but weaker glow from 21:46:30 to 21:48:10 UT (6–16% increase) and the second glow from 21:49:30 to 21:50:30 UT (10%), which both might originate from a static structures we fly in and out of, similar to the structure reported by Kochkin et al. (2017). During the time we observe the increased gamma rays we were flying at constant magnetic latitude so the increase in gamma ray intensity cannot be explained by less magnetic shielding at higher magnetic latitudes. The other gamma ray detector, iSTORM (Figure 2f), also detected the first glow but due to its lower sensitivity did not see the weaker second glow.

Between the two glow regions, the cloud top (Figure 2d) was elevated more than 1 km (close to 15 km) which is higher than before and after (13.5 km), indicating a local region of very strong convection often referred to as "overshooting top."

The black lines in Figures 2a–2c show the three electric field components with 0.1-s resolution, and the spikes are signatures of lightning strokes. The red lines are 20-s running average which can be considered as a rough estimate of at least the direction of the ambient field at 20 km. A 20-s average will filter out the lightning stroke signals that have a ~1-s decay time. Consequently, it can give information about the large-scale dynamics of the strong convective system. The choice of 20-s average is also reasonable given the speed of the aircraft (210 m/s) and the temporal resolution of the LIP instrument (0.1 s). From Figure 2c we see that the 20-s running average of the $E_Z$ component indicates a downward pointing static electric field between the two glow observations but is positive before and after the two glow observations. This is a first indication of an inverted charge structure in the core (negative upper charge layer) and lower positive charge structure beneath and surrounding the core. During the glows $E_Z$ is both positive and negative. We also notice that all the transients (spikes in the black curve) are positive as they have a fast positive rise and a slower decay. This is true both when the running average is positive and negative and can be interpreted as the removal of negative charges from the upper main negative layer. This is another indication of flying over a reversed charge structure and will be explored further in the next section.

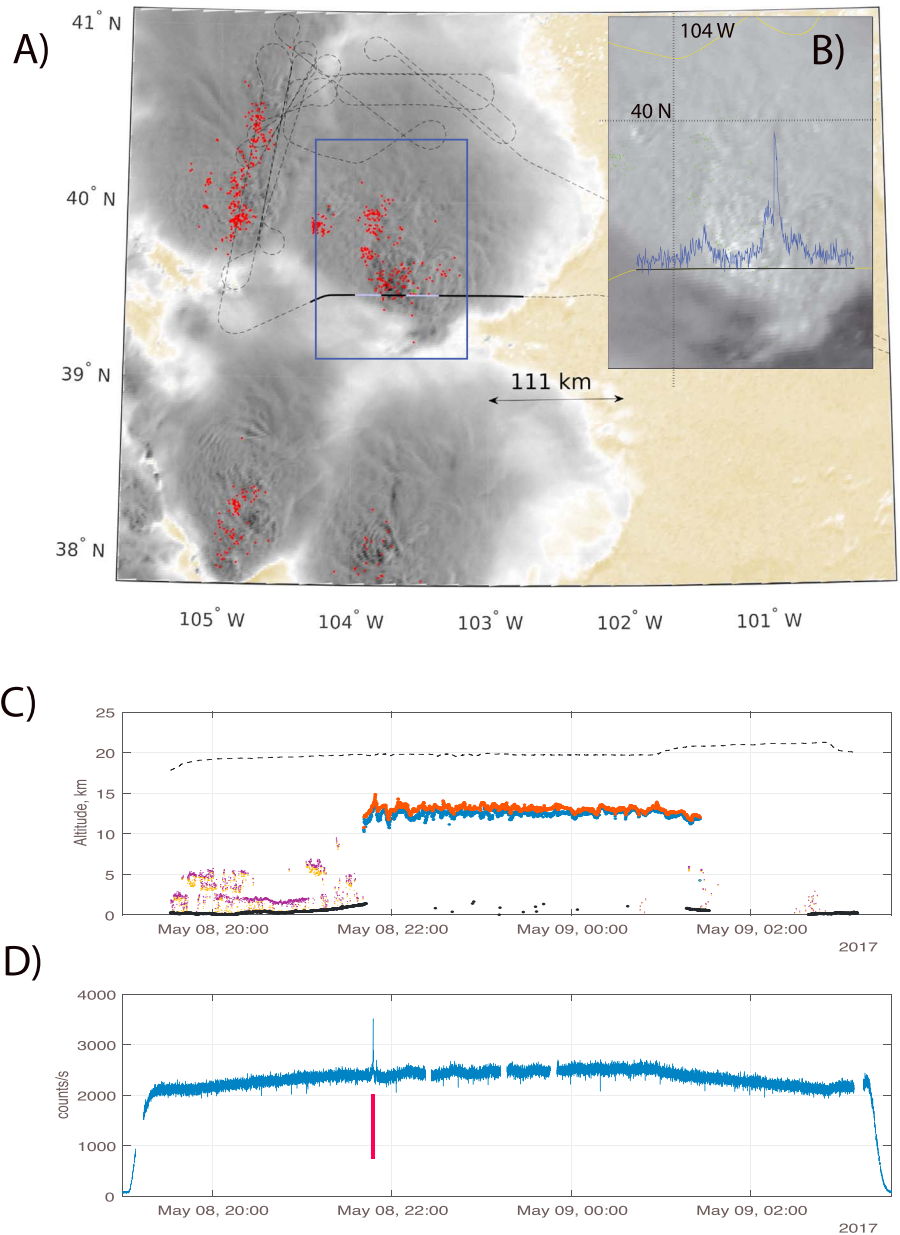

**Figure 1.** (a) Trajectory of ER-2 flying over two convection cells in Colorado just north and northwest of Denver. The cloud images provided by the National Oceanic and Atmospheric Administration GOES is accumulated over 7 min centered around 21:45 UT. Black solid line marks the time of observations studied in this paper, and light gray segments indicate when the gamma glow was observed. No gamma glow was observed during the overpass of the convective system in the northwest. Red dots mark lightning activity detected by WWLLN ±5 min around 21:47 UT. (b) The inset shows the gamma ray observation along the path (time is running to the left). (c) Flight altitude (dashed line) and the upper cloud layer (red and blue) as measured by Cloud Physics Lidar. (d) The gamma ray observations (UIB-BGO >300 keV) during the entire flight on 8 May and the time of glow observation is marked with red vertical line.

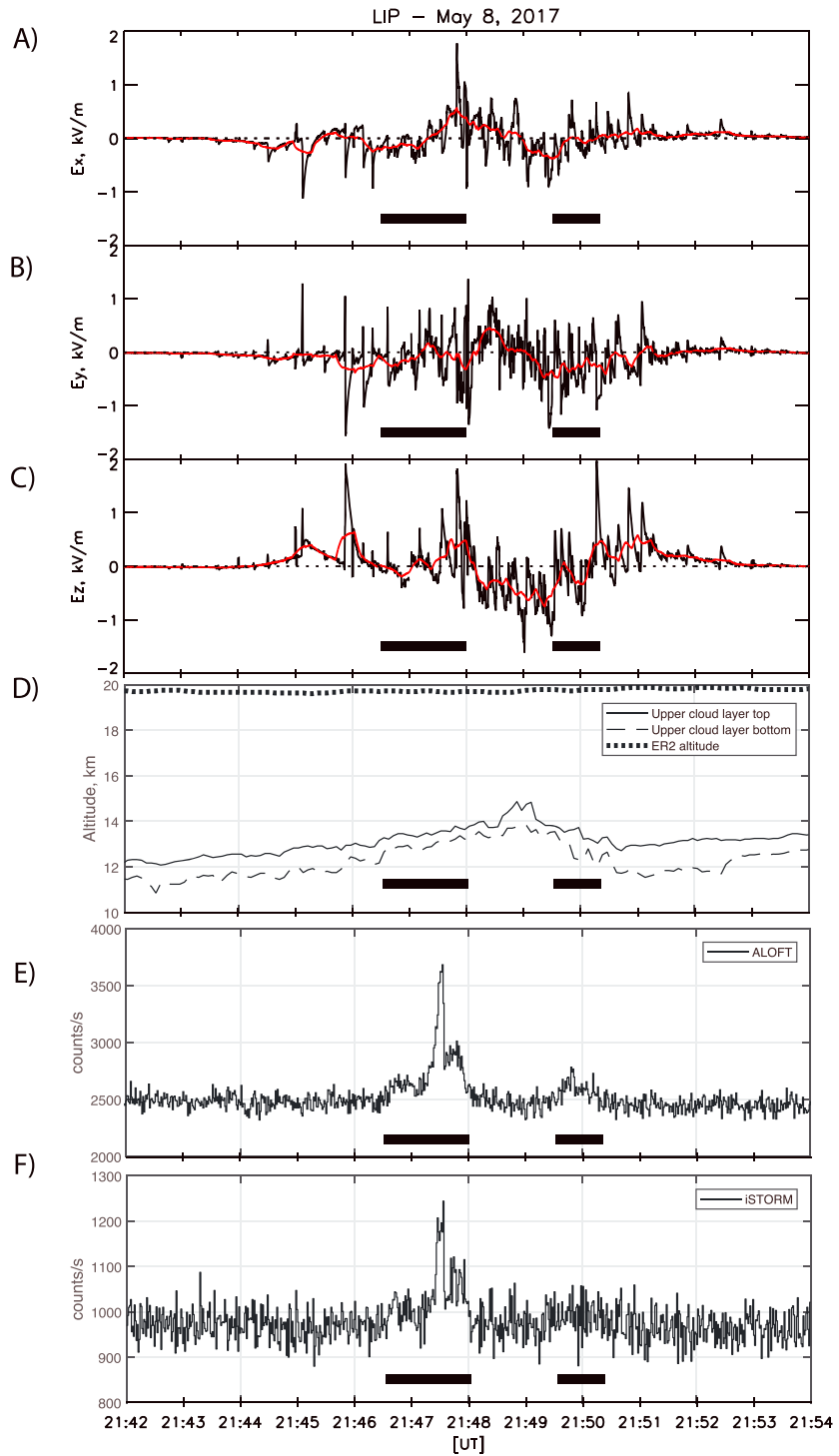

**Figure 2.** Measurement of quasi-static electric field and gamma rays at 20 km. () $E_X$, (b) $E_Y$, and (c) $E_Z$ measured by Lightning Instrument Package with 0.1-s resolution. Red line is a 20-s running average. (d) Flying altitude and cloud top measured by Cloud Physics Lidar (5-s resolution). Two horizontal dashed lines are added to point out the altitude of cloud top during the glow and between the glow. (e) UIB-BGO (>300 keV) and (f) the two large plastic detectors in iSTORM (~100 keV to ~8 MeV), both with 1-s resolution. Time intervals of glow observation are marked with black horizontal lines in all panels. LIP = Lightning Instrument Package; ALOFT = Airborne Lightning Observatory for FEGS and TGFs; iSTORM = in Situ Thunderstorm Observer for Radiation Mechanism.

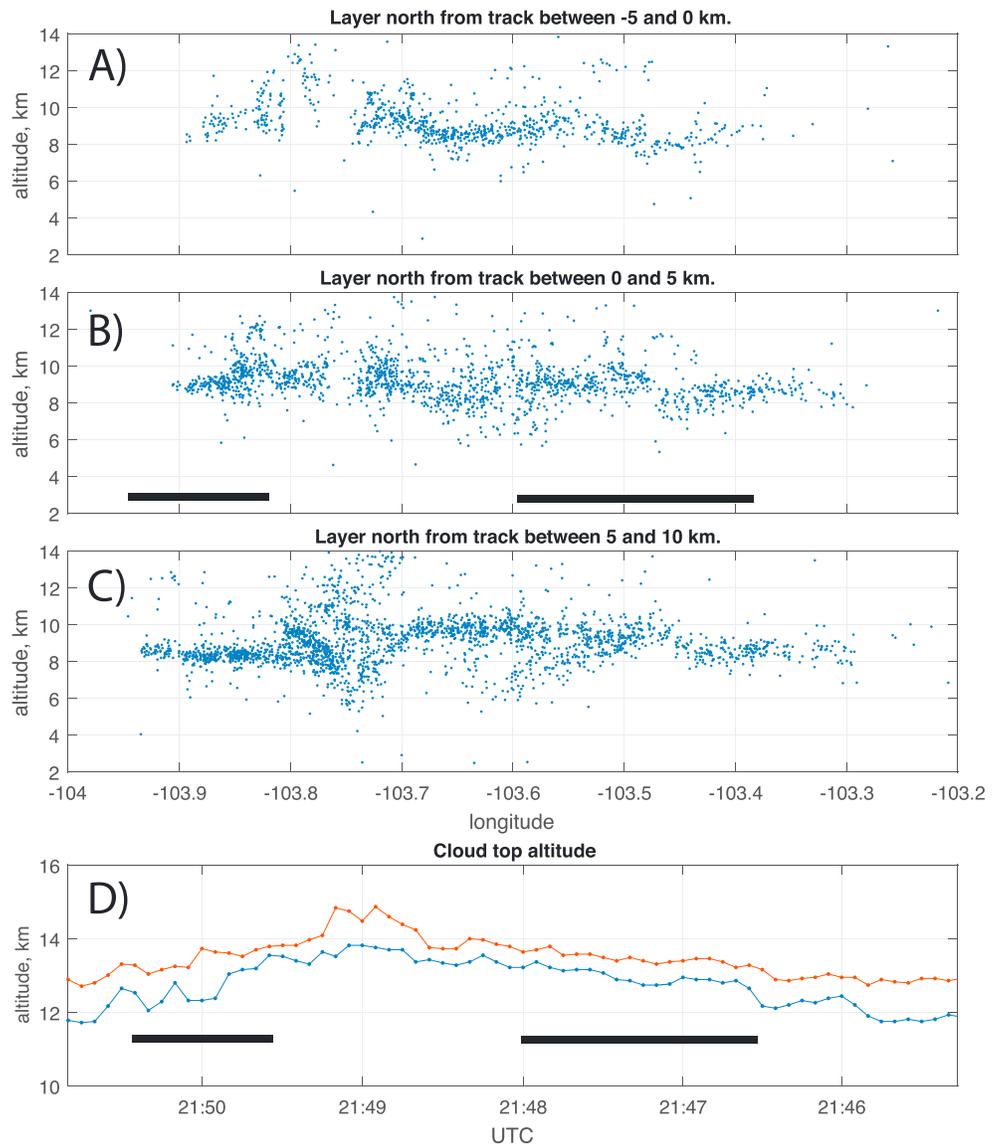

**Figure 3.** Altitude distribution (2–14 km) of lightning activity detected by the Colorado Lightning Mapping Array in selected intervals from flight trajectory ±2 min around the peak of the glow. (a) −5 to 0 km (to the left and south of the trajectory). (b) 0 to 5 km to the right (north) of trajectory. (c) 5 to 10 km (right/north) from trajectory. (d) Cloud tops from Cloud Physics Lidar. Time intervals of glow observations are shown as two thick black horizontal lines between (a) and (b) and in (d).

### 3.2. The Convective Cloud System and Charge Structure

In Figures 3a–3c the altitude distribution of COLMA signals is shown in the ±2-min interval around the peak of the glow (21:47:30 UT) for three distance intervals parallel to the flight trajectory. We will focus on the two upper panels (a and b) that show the activity between the trajectory and ±5 km. From both panels it can be seen that most of the COLMA signals are from the altitude around 8–10 km. We interpret this to be the region of the main positive charge region where negative leaders propagate horizontally. From Figure 3c, which is further to the right (north: 5 to 10 km from the trajectory) and closer to the core of the cloud system, we can see that the activity is spanning an altitude range from 6 to 14 km between −103.8° and −103.7° longitude, which is where the updraft is largest (Figure 3d).

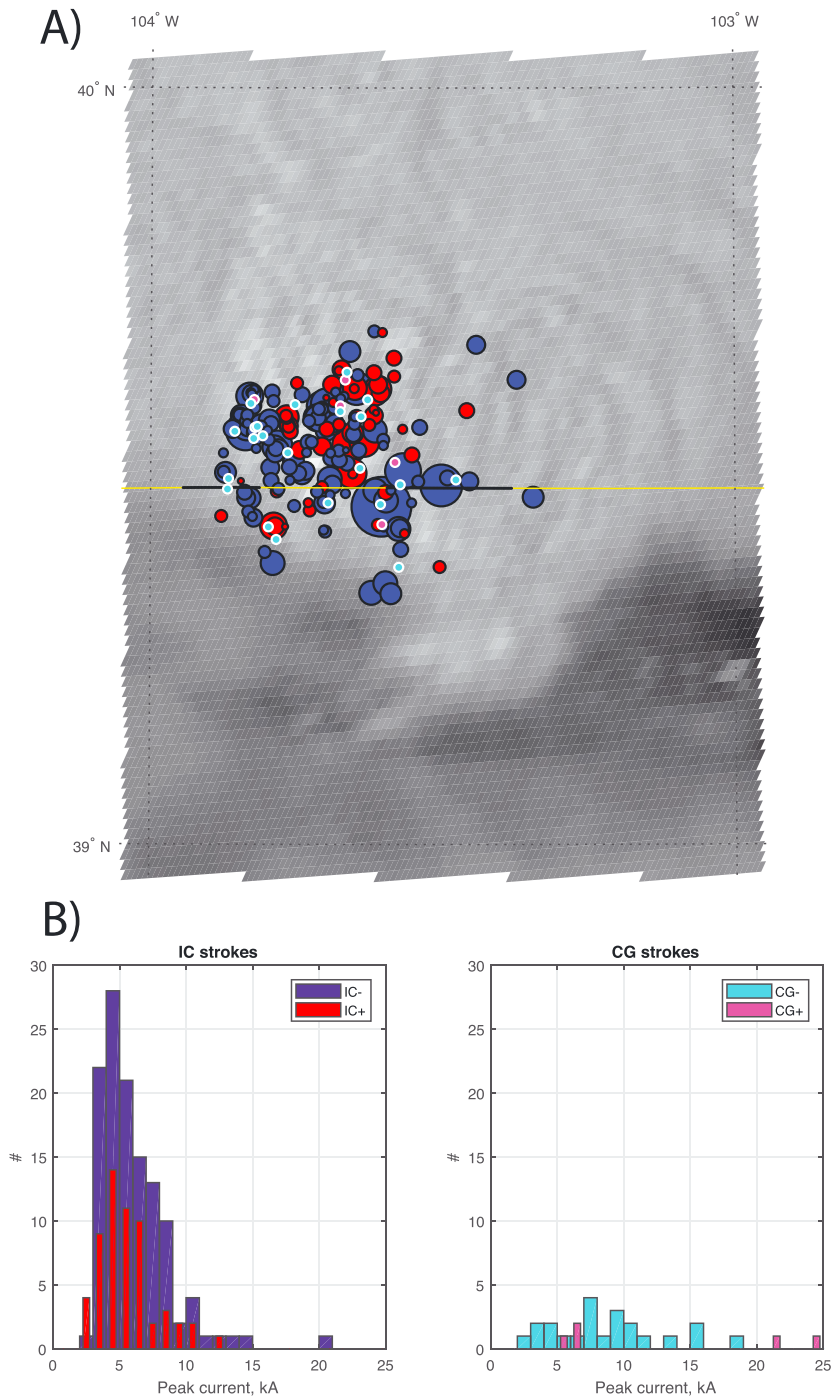

**Figure 4.** (a) The National Lightning Detection Network data ±3 min around 21:49:30 UT spanning the time interval of glow observations showing the polarity and peak currents of intracloud (IC) and polarity of cloud-to-ground (CG) lightning. Blue circles are IC− bringing negative charges downward, and red circles are IC+ bringing negative charges upward. Size of circle of IC indicates the peak current magnitude ranging from 2 to 20 kA. Light blue small dots are CG−, and magenta small dots are CG+. (b) The distribution of peak current magnitudes for IC− and IC+ (left) and for CG− and CG+ (right).

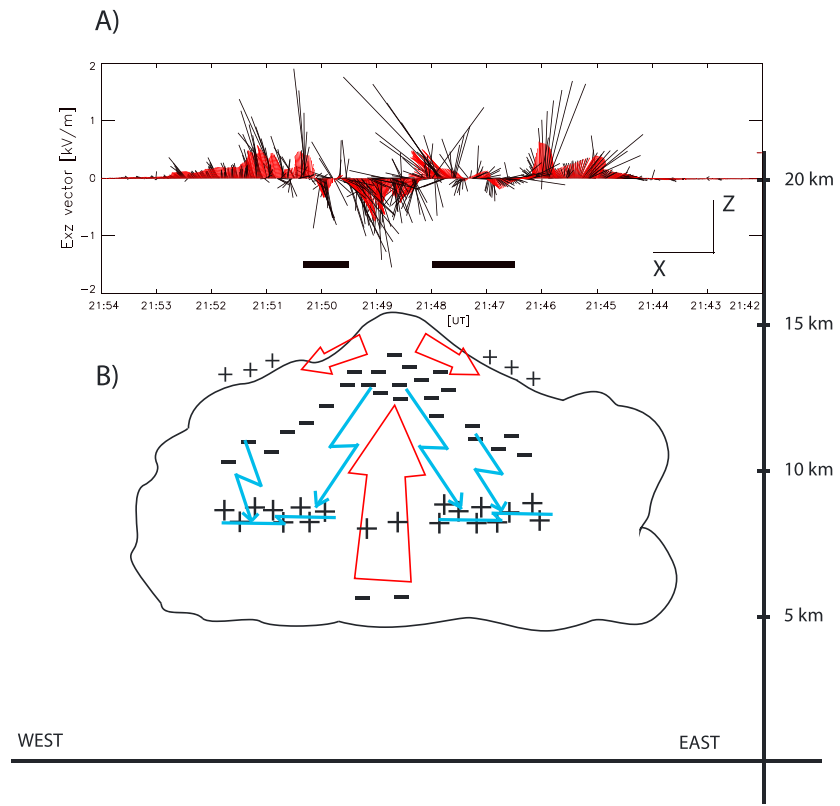

**Figure 5.** (a) The $E_{XZ}$ vectors from Lightning Instrument Package data. Notice that time is running from right to left. Black vectors are every tenth of the 0.1-s resolution data, and red vectors are every tenth of the 20-s running average. (b) A possible charge structure of the cloud in the $XZ$ plane just underneath the aircraft. $X$ and $Z$ is defined in (a). Red large arrows show the strong convection. Minus and plus show the two main negative and positive charge regions, while the small plus indicates the positive screening layer on both sides of the convective core. Blue arrows illustrate the lightning activity.

To further explore the charge structure of the cloud system, we show the NLDN data ±3 min around 21:49:30 UT spanning the time interval of glow observations. In Figure 4a blue (red) circles are IC negative (positive), IC− (IC+), and the size of circle indicates the peak current magnitude. The CG strokes of both polarities are also shown. For the entire cloud system IC− strokes are dominating and even more so along the track of the aircraft. The distribution of peak current magnitudes (Figure 4b) supports this. This means that the dominating charge transfer is to bring negative charges down from the upper negative charge layer. This is an independent indication of an inverted charge structure.

In Figure 5 we suggest a possible charge distribution for this cloud system which is consistent with the LIP electric field measurements, the lightning detection by COLMA and NLDN, and the cloud top altitude from CPL. Figure 5a shows the $E_{XZ}$ vectors from LIP and as the ER-2 was flying to the west time is running to the left. The 20-s running averages (red vectors) indicate a downward field between the glow observations supporting our suggestion of having a main upper negative charge region, as shown in Figure 4b. This is where the main updraft is, and the altitude is consistent with COLMA data in Figure 3c. We have also indicated a fountain effect implying that a positive screening layer above the negative core is not very likely, but such a screening layer could well be on either side of the core. Before and after the glow region, the 20-s running averages of the $E_{XZ}$ vectors (red) are positive pointing away from the cloud structure. Together with the COLMA activity seen in Figures 3a and 3b, we interpret this to be the signatures at 20 km from the main lower positive charge region at about 8–10 km. In the regions where we observe the glow the $E_{XZ}$ vectors are both up and down and we suggest that this is the region where the two main charge regions overlap.

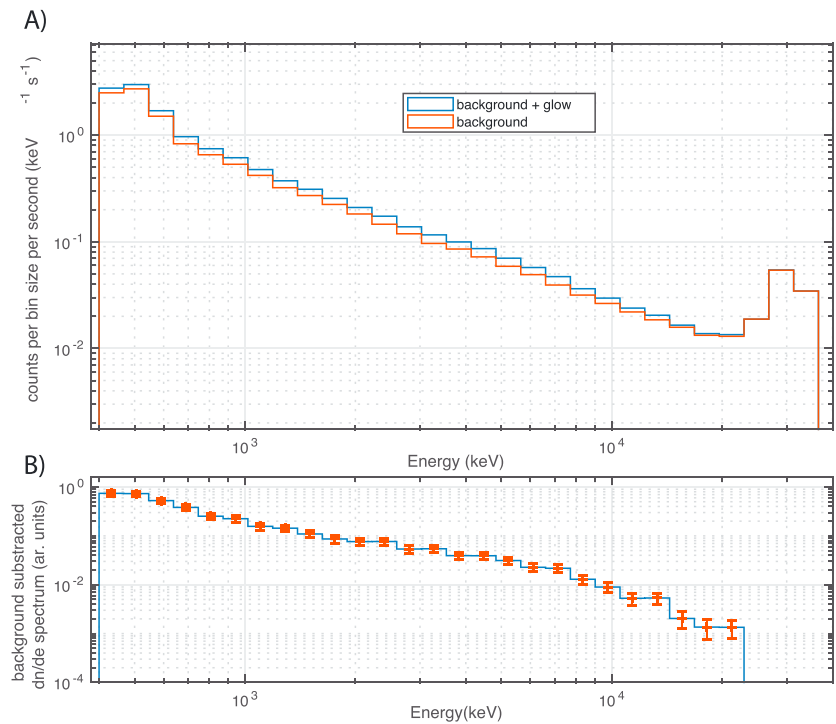

**Figure 6.** The energy spectra observed before and during the glow. (a) Background spectrum before: (21:30–21:38 UT) and during the glow (21:46.30–21:48.10 UT). (b) The background subtracted spectrum during the glow.

We have also indicated that the main lightning activity will neutralize negative charges in the upper charge region, giving positive transients at flight altitude (Figure 2c). The lightning activity with removal of negative charge is over an extended altitude in the core but at lower altitudes before and after (Figure 3). Consequently, it is likely that the main charge layer extends to lower altitudes on either sides of the core, as depicted in Figure 5. We emphasize that this is a simplified sketch of only the main charge regions. There may be mixed regions and pockets of charges in addition to what we suggest here. Nevertheless, as the data from both COLMA and NLDN indicate that the main activity is between the two main charge regions, we will use this charge structure as a reference when we explore possible mechanisms for glow production.

To summarize, all our data indicate that we are passing over an inverted or anomalous charge system. After passing this cloud system, the ER-2 made a right turn and flew northward over the northwest cloud system (Figure 1a). This system was studied by Rutledge et al. (2017) who found this also to be an inverted/anomalous charge system, and they claimed that such structures are common in Colorado close to the Rocky Mountains (S. Rutledge, personal communication, March 15, 2019). The NLDN data from this cloud system (not shown) also support this claim.

## 4. Modeling Glow Production Mechanisms

To explore production mechanisms for the glow, the gamma ray observations give us two important constraints. First, the flux enhancement should be at least 10% for the minute long glows and 45% for the shorter one. Second, observed and modeled spectra should be consistent. In Figure 6 the spectra observed before and during the glow, as well as the background subtracted spectrum, are shown.

We will explore whether the glows are produced locally (above aircraft or between the cloud top and the aircraft) or inside the thundercloud. For these locations we will consider two mechanisms for producing the glow and in both cases we assume that it is the cosmic background flux that is enhanced. The first one is similar to the MOS mechanism suggested by Chilingarian et al. (2012, 2014) for explaining the Thunderstorm Ground Enhancements but here applied to higher altitudes. This idea builds on the ideas suggested earlier by McCarthy and Parks (1992) to explain the first glow observations from the early 1980s (McCarthy & Parks, 1985; Parks et al., 1981). In this case the background flux of all the energies we detect can be enhanced by an electric field above the aircraft or between the aircraft and the cloud top. For the glow production inside the cloud or more exactly between the main charge regions, we will consider the RREA mechanism, first proposed by Gurevich et al. (1992) and further developed and tested by others (e.g., Coleman & Dwyer, 2006; Lehtinen et al., 1999; Skeltved et al., 2014). Although the MOS mechanism can enhance the background flux significantly locally, only RREA can enhance the flux sufficiently to overcome the attenuation from a production location within the cloud up to aircraft altitude at 20 km. The attenuation of gamma rays from a source region within the cloud (8–10 km) to 20 km is a factor of 100–1,000. In addition, the flux of cosmic background at 8–10 km is only 0.4 of the flux at 20 km (not shown but derived from our measurements during the flight descent), which means we need a multiplication of a factor of 250–2,500 to obtain an increase that would be detectable at 20 km. We will explore these two mechanisms using PARMA and Geant4 modeling.

Before the modeling results are presented, we will present a brief back-on-the-envelope estimate of the MOS mechanism that can serve as a reference and validity check for the modeling results. The mathematical derivation is presented in Appendix A.1, and the main points are the following: For an incoming flux of cosmic rays, the electrons (positrons) will be enhanced for upward (downward) electric fields. This is because an external field will accelerate the electrons and increase the attenuation length compared to a zero electric field. At 20 km we find that an electric field of 1.6 kV/m (5.6 kV/m) is required to result in a 10% (45%) increase of the electron flux. These fields correspond to 9% and 31% of the $E_t$ of 17.9 kV/m at 20 km.

### 4.1. Monte Carlo Simulations

To evaluate the glow production scenarios, we have performed Monte Carlo simulations. Two simulation codes were realized using the Geant4 toolkit. Geant4 is developed by the European Organization for Nuclear Research in association with a worldwide collaboration. It is used to simulate particle propagation through matter (Agostinelli et al., 2003; Allison et al., 2006) with or without electromagnetic fields. The ability of Geant4 to accurately simulate particle propagation, acceleration, and all multiplication mechanisms in the context of thunderstorms and high-energy atmospheric radiation was extensively tested by Skeltved et al. (2014), Rutjes et al. (2016), and Sarria et al. (2018).

The initial cosmic ray seed particles considered here (protons, neutrons, photons, electrons, and positrons) are produced using the distributions generated by the PARMA code (Sato et al., 2008), which has a realistic density profile of the atmosphere included. Each particle with its momentum and energy that enters 50-km altitude is then fed into the Geant4 model which propagates the particles further. The altitude of 50 km is chosen because any exponentially decreasing electric field (see section 4.2) above the cloud top would be negligible at this altitude. In Geant4 the atmosphere is simulated between 0- and 50-km altitude, with composition and densities following the NRLMSISE-00 atmospheric model (Picone et al., 2002).

The measured spectrum we try to reproduce is the background subtracted spectrum shown in Figure 6b. To compare this spectrum with a simulated spectrum, a reference (background) simulation was performed, with no electric field ($E = 0$), which is subtracted from the spectra simulated with $E \neq 0$ (the geometry of this field is explained in section 4.2). The variation of particle flux is also expressed with respect to the reference level $E = 0$. The measured and simulated background spectra are consistent with each other within 10% below 25 MeV. Since we are only interested in the relative intensity increase, between having an electric field versus no electric field, this 10% discrepancy should not affect our results.

Two simulation setups were designed for testing the glow production scenarios discussed previously

1. Model A that simulates a large-scale (tens of kilometers) electric field starting from the top of the cloud (assumed at 13 km) toward higher altitudes (50 km).
2. Model B that simulates a smaller-scale (full size of 2 km) electric field inside the cloud with center location from 9 to 12 km.

Models A and B have different sets of free parameters. In both cases, all the particles that cross 20-km altitude (where the aircraft is located) are recorded. The fluxes of photons, electrons, and positrons are saved as well as their energy spectra. The photon energy spectrum is then passed through the detector response matrix of the UIB-BGO detector.

The relative distribution of photons, electrons, and positrons (>300 keV) at 20-km altitude is ∼91.7%, ∼6.0%, and ∼2.3% (Model A), while at 12 km it is ∼92.7%, ∼5.3%, and ∼2.0% (Model B), respectively. In both cases, photons are largely dominating.

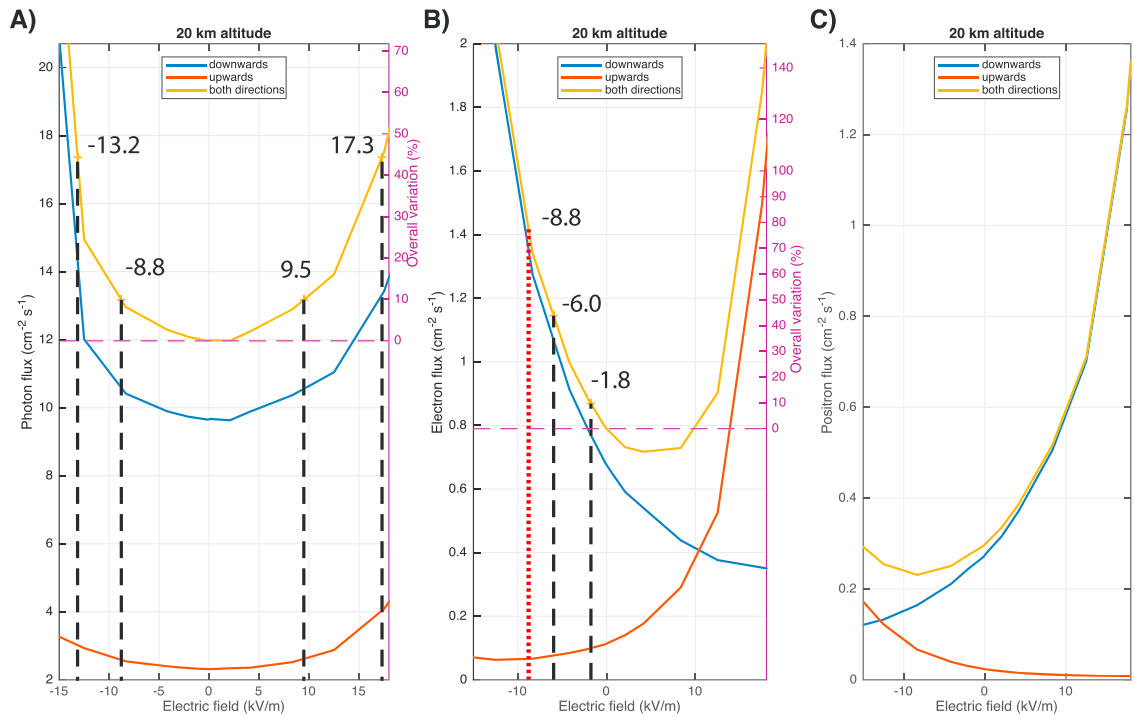

**Figure 7.** Variation of the flux at 20-km altitude as function of the electric field at the same altitude ($E_{ac}$) for (a) photons, (b) electrons, and (c) positrons. The different flux scales on the left side of the panels reflect the relative number of photons (~91.7%), electrons (~6.0%), and positrons (~2.3%) above 300 keV for the background values ($E_{ac} = 0$). Unit on the right side is the increase (%) relative to the background values (when $E_{ac} = 0$). Positive (negative) is a downward (upward) electric field at 20 km ($E_{ac}$). Blue (red) lines are particles entering 20 km from above (from below), while yellow lines are all particles passing 20 km. The dashed vertical gray lines (panels a and b) indicate the 10% and 45% flux increases. The red dashed vertical line in panel (b) is electric field strength needed for increasing the total photon flux by 10% (see text for explanation). The photon flux (panel a) increases by 10% for $E_{ac} = -8.8$ kV/m and $E_{ac} = 9.5$ kV/m and by 45% for $E_{ac} = -13.2$ kV/m and for $E_{ac} = 17.3$ kV/m. Increases of the electron flux (panel b) by 10% and 45% are obtained for field strengths of $-1.8$ and $-6.0$ kV/m, respectively.

Furthermore, the photons are also more efficiently detected as they can penetrate more easily the different materials in front of the UIB-BGO detectors, for example, fuselage, instrument covers, and detector housing. Thus, it is reasonable to only record the photons for the comparison between measurement and simulations.

There are two observational constraints for finding the best fit between the simulations and the measurements: (1) the increase of the detected photon flux above background should be around 10% to 45% and (2) the shape of the measured energy spectrum (Figure 6b), which was binned using 26 bins. The goodness of fit between the simulations and the measurement is expressed by the reduced chi-square ($\chi^2_{\text{red}}$). A value close to 1 indicates a good fit. Depending on the chosen acceptance level and the number of degrees of freedom, a critical value of $\chi^2_{\text{red},c}$ exists above which the fit is considered incompatible.

### 4.2. Glow Produced Above the Thundercloud (Model A)

The cloud top is assumed to be at $h_{\text{top}} = 13$-km altitude, where the electric field magnitude is $E_{\text{top}}$. The absolute value of the electric field $E(h)$ decreases exponentially as function of increasing altitude, with a scale height $H = 8$ km (Rakov & Uman, 2003)

$$E(h) = E_{\text{top}} \exp\left(\frac{-(h - h_{\text{top}})}{H}\right) \quad (1)$$

This exponentially decreasing field with altitude (13–50 km) is implemented over a horizontal area of 40 km × 40 km. $E_{\text{top}}$ is the only free parameter of Model A. It was tested from $-42$ to $+42$ kV/m. A positive electric field is a downward field and implies that electrons are accelerated toward higher altitudes and positrons in the opposite direction. The electric field at the aircraft altitude ($E_{ac}$) is directly related to $E_{\text{top}}$ (equation (1)): $E_{ac} = 0.417 \, E_{\text{top}}$ or $\pm 17.5$ kV/m as used in Figure 7.

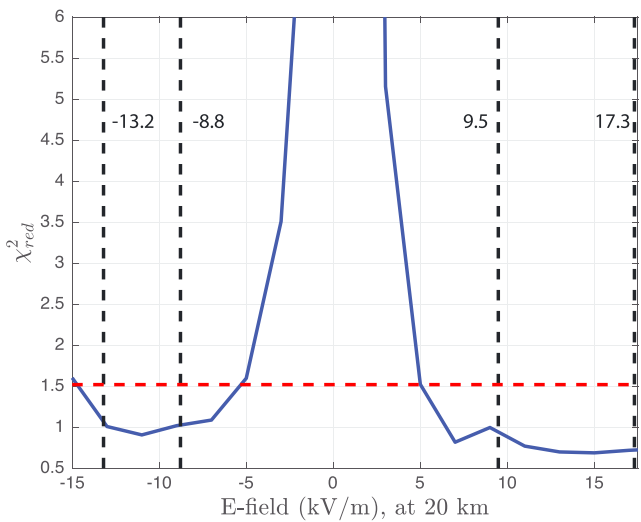

**Figure 8.** Reduced $\chi^2$ values of the simulated photon spectra fits to the measured spectrum. The critical value $\chi^2_{red,c} = 1.52$ assuming a 5% acceptance level is highlighted by a dashed red line. A $\chi^2_{red}$ value below $\chi^2_{red,c}$ indicates a compatible fit. The required field strengths for 10% and 45% increase of photon flux are marked by black dashed lines for both downward (positive) and upward (negative) fields.

Figure 7 shows the modeling results for photons, electrons, and positrons at 20 km for downward (positive) and upward (negative) electric fields. The flux units on the left side in all panels reflect the relative number of photons (∼91.7%), electrons (∼6.0%), and positrons (∼2.3%) in the cosmic background at 20 km when $E_{ac} = 0$. In Figure 7b we see that for electrons entering from above (blue line) the flux increases for upward (negative) fields and is reduced for downward (positive) field, as expected. A similar behavior, but opposite, is seen for positrons. For electrons entering from below (red line) relatively large field strength is required for making the electrons turn around and increase the upward flux of electrons in the downward field below 20 km. To obtain an increase of electron flux of 10% and 45%, upward fields ($E_{ac}$) of −1.8 and −6.0 kV/m are required, which correspond to 10% and 34% of the local $E_t$. We emphasize that this is only valid for increasing the flux of electrons and can then be compared with the back-on-the-envelope estimates of −1.6 and −5.6 kV/m. They are both in fairly good agreement and indicate that the assumptions made in Appendix A.1 are reasonable.

However, for the total flux of photons much larger field strengths are needed to obtain the observed increases. For a 10% increase $E_{ac} = -8.8$ and $E_{ac} = 9.5$ kV/m are required (49% and 53% of $E_t$) and for a 45% increase the required electric field strengths are $E_{ac} = -13.2$ kV/m and for $E_{ac} = 17.3$ kV/m, 74% and 97% of $E_t$. The values are slightly larger for positive (downward) fields as the photon increase in this case is due to electrons that are turned around by the electric field and then interact with air to produce photons that enter from below. All these ambient field strengths are an order of magnitude larger than what we derive (20-s running average) from our electric field measurements.

As pointed out above, our instrument mainly detects photons, which are the photons from the incoming cosmic background (which dominates ∼91.7%) and photons produced by the enhanced electron and positron flux when interacting with air. The contribution of photons produced by electrons and positrons interacting with aircraft and shielding material was tested and found negligible, below 1% on the overall spectrum and below 5% for the energy bin containing 511 keV. In Figure 7b we have indicated with red vertical dashed line the electric field strength ($E_{ac} = -8.8$ kV/m) that are needed to give a 10% increase in photon flux (Figure 7a). We find that an 80% increase in electron flux is needed because electrons are only a small fraction of the cosmic background (∼6.0%). This implies (see Appendix A.2) that each electron must produce on average ∼2 photons in interaction with air to give a 10% increase in photon flux.

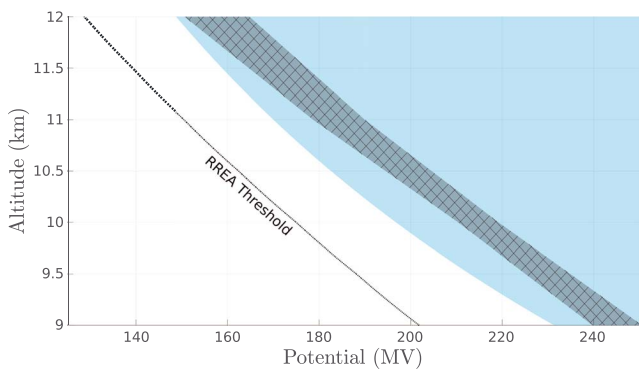

**Figure 9.** Model B simulation results of acceptable $\chi^2_{red}$ values for both the spectral fit and the enhanced fluxes as function of potential (ΔU) and altitude (H). The blue area indicates the altitude-potential range where the reduced $\chi^2_{red}$ of the spectral fit of the simulation to the measurement is below 1.53, which is the critical value for the 5% level for 23° of freedom. The gray hatched area denotes the altitude-potential range where the photon flux increase (at 20-km altitude and between 300 and 30 MeV) with respect to background is between 10% and 45%. A constant electric field over 2 km is assumed, and the values are shown for the middle of this region; for example, 11 km is for an electric field from 10 to 12 km. Black line is the RREA threshold. RREA = Relativistic Runaway Electron Avalanche.

To compare measured and modeled spectra, the modeling results from Figure 7 are passed through the energy response matrix of the detector. The quality of the photon spectrum fits with the measurements, as function of $E_{ac}$, is presented in Figure 8. Since we have 24° of freedom, and assuming a standard 5% acceptance level, the critical $\chi^2_c$ is 36.4 or equivalently $\chi^2_{red,c} = 1.52$. Model fits with $\chi^2_{red}$ values below this threshold are considered compatible. Figure 8 shows that for positive fields the compatible values are for $E_{ac} > 5$ kV/m and goes up to above 17.3 kV/m. For negative fields, the compatible values are for $-15 kV/m < E_{ac} < -5 kV/m$. Thus, the required field strengths (marked with black dashed lines in Figure 8) for obtaining 10% and 45% increase in photon flux for both downward (positive) and upward (negative) electric fields are compatible solutions.

In Figure 10a we present the energy spectra for a 10% increase of photon flux for both downward (+9.5 kV/m) and upward (−8.8 kV/m) electric field strength at 20 km (both around 50% of $E_t$). Although there is a discrepancy in the 3- to 7-MeV range, both spectra are acceptable solutions based on their $\chi^2_{red}$ values.

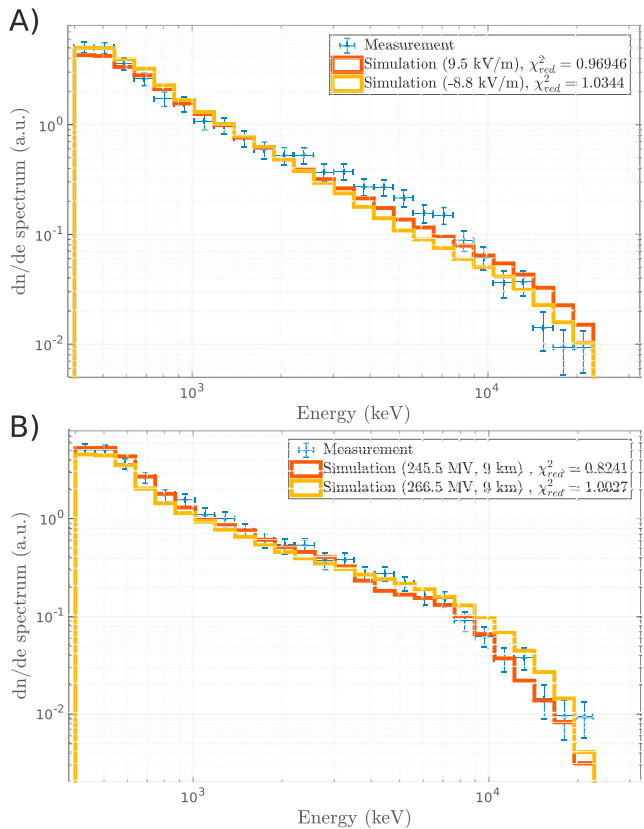

**Figure 10.** Spectrum comparison between the measurement (background subtracted) and fits obtained from Geant4 modeling. (a) Model A photon spectra for electric field strengths required at 20 km for a 10% increase in photon flux. (b) Model B photon spectra for two potential values between 8 and 10 km. LIP = Lightning Instrument Package; ALOFT = Airborne Lightning Observatory for FEGS and TGFs; iSTORM = in Situ Thunderstorm Observer for Radiation Mechanism.

### 4.3. Glow Produced Inside the Thundercloud (Model B)

For a production mechanism inside the cloud, we will only show the results from RREA, as the MOS mechanism inside the cloud is not able to produce the enhancement needed to overcome the attenuation from <13 to 20 km. The relativistic feedback mechanism is taken into account by our simulations, as it is included in Geant4 by default (see, e.g., Skeltved et al., 2014). It refers to the possibility of positrons and backscattering X-rays to come back inside the avalanche region and consequently seed more RREAs. All the simulations we present here are in the non-self-sustaining feedback regime ($\gamma < 1$), because the used electric fields are not strong enough. For more information about the relativistic feedback theory, see section 2.6 of Dwyer et al. (2012). In Model B we assume a constant vertical electric field located over a 2-km altitude range (between the main charge structures). There are two free parameters:

- The altitude of the center of the constant electric field region, noted $H$.
- The applied potential difference, noted $\Delta U$. A positive potential implies that electrons are accelerated upward and positrons downward.

Figure 9 presents the simulations results for both the spectral fits and the particle flux increases. It shows the results for positive potentials (corresponding to a normal thunderstorm), while the results with negative potentials are discussed below. In Figure 9 the gray hatched area highlights the altitude-potential region where the photon increase is between 10% and 45% and the blue area indicates the area of the $\chi^2_{red}$ of the simulated spectra that fits with the measurement. Since there are 23° of freedom and assuming a standard 5% acceptance level, the critical $\chi^2_c = 35.2$ or equivalently $\chi^2_{red,c} = 1.53$.

For a vertical downward electric field the cases with $\Delta U$ from 150 MV (electric field between 11 and 13 km) up to 240 MV (electric field between 8 and 10 km), both about 1.2 times the $E_t$, are compatible with the measurements. In Figure 10b we present two valid spectrum fits from Model B for positive potential.

The presented results of Model B are for a positive potential that corresponds to a normal polarity thundercloud, where the electrons are accelerated toward higher altitudes. However, in section 3.2 we found strong support for having an anomalous polarity storm, which would correspond to a negative potential in this modeling. The simulation with negative potentials was also performed and could not produce any acceptable spectrum fit, due to the prominence of the photons from annihilation, as a massive amount of positrons are accelerated toward higher altitudes. The photon flux at 20 km could barely reach a 10% increase that only happened for $H > 13$ km (which is very close to the cloud top) and $|\Delta U| > 120$ MV.

## 5. Discussion

We have presented the gamma ray glows observed by two detectors at 20 km, inferred a possible charge structure of the thundercloud system from in situ electric field measurements and ground-based lightning detection networks. We have explored two different production mechanisms by combining Geant4 and PARMA modeling assuming glow production above the cloud (13–50 km) and within the cloud (9–12 km).

### 5.1. Cloud Charge Structure

We have inferred that the ER-2 aircraft was flying over an inverted charge structure when the glows were observed. This is based on both the in situ electric field measurements (Figure 2c) that show that the lightning strokes have a fast positive rise time and slow decay, indicating that negative charges move downward. When applying a 20-s running average to the $E_{XZ}$ component (Figure 5a) we get a downward ambient field when we pass close to the cloud top, also suggesting negative charges at the cloud top. Furthermore, the altitude distribution of COLMA signals indicates that the main positive charge region is at 8–10 km, well below the cloud top at 13–15 km. Finally, the NLDN data (Figure 4) support the claim that IC− are dominating in the entire cloud system and even more so just underneath the aircraft trajectory. We are therefore fairly confident that ER-2 is passing over an inverted charge structure and that the large-scale electric field inside the cloud is most likely an upward field.

Above the cloud the field orientation is more complicated. From the running averages of the electric field measurements (Figure 5a) we cannot tell whether the electric field is upward or downward at 20 km during the glow observations. It is upward before and after the glow and downward between the glows but is both upward and downward field during the glow. One could also question whether the 20-s running average is a too simplistic way to estimate the ambient field, and it may be that we do not have the magnitude of the ambient field correct.

### 5.2. Glow Production Mechanism

When describing the glow observations, we have pointed out that there are two enhancements of gamma ray of about 10% that lasted for about 1.5 and 1 min and a shorter (15 s) but larger (45%) enhancement. A reasonable interpretation of the 1.5- and 1-min-long glows is that we fly over a steady glow production region of about 19 and 13 km (aircraft speed is 210 m/s), respectively. The 45% increase with a rise time of about 15 s seems too long to be a transient and is probably a localized region of enhanced electric field. However, we are not able to distinguish between temporal and spatial changes.

We have tested two production mechanisms for the glow production assuming different locations.

For a local production (below or above 20 km) we have only considered the MOS mechanism, as we find it unlikely that the field strength at this altitude is above the RREA threshold. The modeling results of the MOS mechanism show that both upward and downward fields can produce the enhancements we observe (10% and 45%) as well as the spectral shape. However, the required field strengths are about a factor of 10 larger than we derive from the 20-s running average. As the latter method of estimating the ambient field could be questioned, we will not rule out the MOS as a possible mechanism for producing the glows.

For a production mechanism inside the cloud we also find solutions that match both the enhancements and the spectral shape but only for downward field. If this is the case, it must be some large region that does not have an inverted charge structure, maybe in the region where there are IC+ lightning strokes (Figure 4a) north of where the aircraft flies. For this to be true the source of the glow is not a spatial structure but a temporal mechanism that is turned on-off-on-off. Furthermore, from Figure 9 one can see that rather large potentials are needed to match the measurements, which means that the large-scale ambient electric field has to be ∼1.2 times larger than the local RREA threshold, $E_t$.

This raises the question if such a large field can be sustained for a long time (1 min) and over large regions (>12 km) without the RREA discharging the cloud system. Following a similar approach as in Kelley et al. (2015; see Appendix A.3 for details) we find that the multiplication factor is too low for self-sustained feedback mechanism and that the discharge current is low enough to be compensated by the convection charging current and the RREA will not discharge the cloud. However, this raises another question. Why do we not see any terrestrial gamma ray flashes? There is a lot of lightning activity during the glow, and this implies that the electric field will be even higher ahead of the tip of the leaders. Since this scenario assumes a downward field and a region of IC+ lightning, it is likely that the combination of an electric field already above the RREA threshold and the strong field ahead of the leader would lead to the production of an upward terrestrial gamma ray flash (TGF), as recently suggested by Smith et al. (2018). The streamer zone ahead of the leader would also provide sufficient seed electrons for RREA to account for the typical number of relativistic electrons in a TGF. However, we do not observe any terrestrial gamma ray flash.

Since the RREA process requires an opposite charge structure than we see, a large time-varying potential and despite lots of lightning activity that we do not see any TGFs, we conclude that it is less likely that the RREA process is responsible for the observed glows.

## 6. Summary

In this paper we have reported the first glow observations from 20-km altitude and our main findings are as follows.

1. The glows are observed over an inverted charge structure.
2. A possible production mechanism is the enhancement of cosmic background (MOS) by an electric field with field strength well below the RREA threshold. However, for this to be true, we have to assume that the ambient electric field at 20 km is underestimated by an order of magnitude.
3. The RREA model was also tested but requires a normal charge structure contrary to what the data suggest as well as a potential ~1.2 times the RREA threshold.

## Appendix A

### A.1. Estimate of the Electron Flux Increase by the MOS Mechanism

Here we present a brief back-on-the-envelope estimate of the MOS mechanism which can serve as a reference and validity check for the modeling results. These calculations show the conditions on electric field $E$, necessary for the increase in both cosmic ray electrons and the gamma photons produced by these electrons (which is proportional to electron number). If we have a constant supply of relativistic electrons, $S$, due to higher-energy (above the critical energy 80 MeV) electromagnetic particles and hadron primaries and their time rate of absorption, $\nu$, then the electron number $N$ (density, spectral density, or a similar quantity) may be described by equation

$$\frac{dN}{dt} = S - \nu N \tag{A1}$$

The absorption rate is given by

$$\nu = \frac{c}{\lambda} \tag{A2}$$

where $c$ is the velocity of relativistic electrons (close to the speed of light) and $\lambda$ is the attenuation length, which is given by the formula for the RREA multiplication length but with the opposite sign (because $E < E_t$)

$$\lambda = \frac{7,000 \text{kV}}{E_t - E} \tag{A3}$$

where $E_t$ is the RREA threshold of ~280 kV/m (e.g., Coleman & Dwyer, 2006) at sea level and at other altitudes proportional to atmosphere density. $E$ is the external uniform electric field ($< E_t$). Since the formula for density time evolution (equation (A1)) does not take into account spatial dependence of $E$, it is only valid for larger systems when the extension of the field $E$ is $\gg \lambda$. The typical values for $\frac{1}{\nu}$ are seen to be $\ll 1$ ms; thus, the equilibrium is established much quicker than the typical scale of the observed gamma ray glow, and for equilibrium the left side of equation (A1) can be set to zero. The number of relativistic electrons is thus given as

$$N(E) = \frac{S}{\nu(E)} \sim \frac{1}{E_t - E} \tag{A4}$$

and the enhancement with respect to the background with zero field is

$$r = \frac{N(E)}{N(0)} = \frac{E_t}{E_t - E} \tag{A5}$$

and is only valid when $E < E_t$.

$$E = (1 - \frac{1}{r})E_t \tag{A6}$$

If we assume a local enhancement (or close by), the RREA threshold field $E_t$ at 20 km scaled with the atmospheric density given by NRLMSISE-00 (Picone et al., 2002) is ~17.9 kV/m ($\frac{n_0}{n_{20}} = 15.6$). Using equation (A6) with $r = 1.45$ (the 45% increase) $E$ needs to be 5.6 kV/m (at 20 km, i.e., over a few attenuation lengths). For a 10% increase only 1.6 kV/m is needed.

With conductivity following $\sim e^{z/H}$ the field should also decrease with $\sim e^{z/H}$ (where $z$ is the altitude). With $H = 8$ km (pg 9 ; Rakov & Uman, 2003) the field at the cloud top at 13 km should be 13.3 and 3.9 kV/m to give 45% and 10% enhancement, respectively.

Given that equation (A3) is valid for these field strengths the attenuation length at 20 km for electrons is 700 and 550 m. As both $E$ and $E_t$ fall off exponentially the attenuation length at 28 km is not more than 2 km, and we would have more than six attenuation lengths between 20 and 28 km, which satisfies the assumption that the extension of $E$ is $\gg \lambda$.

It should be noted that we have only considered the increase of electrons in this calculation and that this only applies for an upward field. However, a similar estimate could be done for positrons in a downward field.

### A.2. Number of Photons Produced by the Enhanced Flux of Electrons

As presented before, there are about 91.7% photons, 6.0% electrons, and 2.3% positrons (>300 keV) in the background flux at 20-km altitude (i.e., when $E = 0$). We can convert these percentages to arbitrary units of particle number to obtain

$$N_\gamma = 91.7\,a.u. \text{ (photons)}$$
$$N_{e^-} = 6.0\,a.u. \text{ (electrons)}$$
$$N_{e^+} = 2.3\,a.u. \text{ (positrons)}$$

Then we define the quantity $\omega$, the average number of photons produced by each electron (or positron) that are able to cross the 20-km altitude layer. If we assume $\omega \approx 2.5$, then the number of photons reaching this altitude is

$$N_\gamma + N_{e^-} \times \omega + N_{e^+} \times \omega = 112.5\,a.u.$$

Then, by applying an electric field of $-8.8$ kV/m, we know from Figure 7 that the positron number is almost unchanged and that the electron number is increased by $\approx 80\%$ (i.e., ×1.8). Therefore, the number of photons reaching 20-km altitude is

$$N_\gamma + 1.8 \times N_{e^-} \times \omega + N_{e^+} \times \omega = 124.5\,a.u.$$

This corresponds to a 10.7 % variation with respect to when $E = 0$ kV/m, which is consistent with the photon flux variation shown in Figure 7.

We can also calculate the value of $\omega$ that gives exactly a 10% variation with

$$\omega = \frac{0.1 N_\gamma}{(0.7 N_{e^-} - 0.1 N_{e^+})} \approx 2$$

### A.3. Hypothesis of RREA Inside the Cloud

We follow reasoning of Kelley et al. (2015). The difference is that we take the seed source of >300-keV electrons for RREA to be $S = 1$ cm$^2$/s $= 10^4$ m$^2$/s instead of 0.25 cm$^2$/s (>1 MeV) used by Kelley et al. (2015). The cloud discharging current is due to both lightning and the effects of runaway electrons. Out of the latter, the most important contribution is the ion current, because ions created in the RREA process have relatively long lifetime and accumulate during such a long process as gamma ray glow. This current is given by equation (3) in Kelley et al. (2015), namely, $J = 2e\alpha \lambda_{cl} F_{re}$, where we also use their notations. In this equation, $F_{re}$ is the flux of runaway electrons, $\alpha = 8,350$ N/m is the ionization per unit length created by them, and $\lambda_{cl} = 10$–30 m is the estimate of the mean-free path of ions before they attach to water or ice particles (we take a range of values around Kelley et al., 2015, value of 27 m). The relative air density (in respect to standard conditions) is $N = 0.1$–$0.2$ in the altitude range of interest (10–15 km). The initial seed source is multiplied by a factor of $K = 300$–$1,000$ as a result of avalanche (which is well below the values required for the photon/positron feedback mechanism), so that $F_{re} = SK = 3 \times 10^6$–$10^7$ m$^2$/s. Substituting all the values, we get the ballpark estimate range $J \sim 8$–$160$ nA/m$^2$ which is, as in Kelley et al. (2015) case, comparable with other thunderstorm discharging currents and thus may be compensated by the charging currents due to convection. For the horizontal cloud size of ~10 km, that is, area of $\sim 10^8$ m$^2$, the total current is $I \lesssim 20$ A.

To summarize, the considered conditions allow continuous RREA because (1) the multiplication factor is too low for feedback mechanism and (2) the discharging current is low enough to be compensated by the convection charging current.

**Acknowledgments**
The data from the aircraft campaign which are described in this paper are uploaded as supporting information. This study was supported by the European Research Council under the European Unions Seventh Framework Programme (FP7/2007-2013)/ERC grant agreement 320839 and the Research Council of Norway under contracts 208028/F50 and 223252/F50 (CoE). We thank the WWLLN and COLMA networks for the use of their data. We thank Vaisala, Inc. for access to the U.S. National Lightning Detection Network data used in this study. The simulations were performed on resources provided by UNINETT Sigma2—the National Infrastructure for High Performance Computing and Data Storage in Norway, under project NN9526K.

## References

Agostinelli, S., Allison, J., Amako, K., Apostolakis, J., Araujo, H., Arce, P., et al. G EANT4 Collaboration (2003). GEANT4—A simulation toolkit. *Nuclear Instruments and Methods in Physics Research A*, *506*, 250–303. https://doi.org/10.1016/S0168-9002(03)01368-8

Allison, J., Amako, K., Apostolakis, J., Araujo, H., Dubois, P. A., Asai, M., et al. (2006). Geant4 developments and applications. *IEEE Transactions on Nuclear Science*, *53*, 270–278. https://doi.org/10.1109/TNS.2006.869826

Bateman, M. G., Stewart, M., Blakeslee, R., Podgorny, S., Christian, H., Mach, D., et al. (2007). A low noise, microprocessor-controlled, internally digitizing rotating-vane electric field mill for airborne platforms. *Journal of Atmospheric and Oceanic Technology*, *24*, 1245–1255. https://doi.org/10.1175/JTECH2039.1

Chilingarian, A. (2014). Thunderstorm ground enhancements—Model and relation to lightning flashes. *Journal of Atmospheric and Terrestrial Physics*, *107*, 68–76.

Chilingarian, A., Amilyan, B., & Vanyan, L. (2012). Recovering of the energy spectra of electrons and gamma rays coming from thunderclouds. *Atmospheric Research*, *114–115*, 1–16. https://doi.org/10.1016/j.atmosres.2012.05.008

Chilingarian, A., Hovsepyan, G., & Hovhannisyan, A. (2011). Particle bursts from thunderclouds: Natural particle accelerators above our heads. *Physical Review D*, *83*, 062001. https://doi.org/10.1103/PhysRevD.83.062001

Chilingarian, A., Hovsepyan, G., & Vanyan, L. (2014). On the origin of the particle fluxes from the thunderclouds: Energy spectra analysis. *Letters Journal Exploring the Frontiers of Physics*, *106*, 59001. https://doi.org/10.1209/0295-5075/106/59001

Coleman, L. M., & Dwyer, J. R. (2006). Propagation speed of runaway electron avalanches. *Geophysical Research Letters*, *33*, L11810. https://doi.org/10.1029/2006GL025863

Cramer, A. S., Mailyan, B. G., Celestin, S., & Dwyer, J. R. (2017). A simulation study on the electric field spectral dependence of thunderstorm ground enhancements and gamma ray glows. *Journal of Geophysical Research: Atmospheres*, *122*, 4763–4772. https://doi.org/10.1002/2016JD026422

Dwyer, J. R., Grefenstette, B. W., & Smith, D. M. (2008). High-energy electron beams launched into space by thunderstorms. *Geophysical Research Letters*, *35*, L02815. https://doi.org/10.1029/2007GL032430

Dwyer, J. R., Rassoul, H. K., Al-Dayeh, M., Caraway, L., Chrest, A., Wright, B., et al. (2005). X-ray bursts associated with leader steps in cloud-to-ground lightning. *Geophysical Research Letters*, *32*, L01803. https://doi.org/10.1029/2004GL021782

Dwyer, J. R., Rassoul, H. K., Al-Dayeh, M., Caraway, L., Wright, B., Chrest, A., et al. (2004). A ground level gamma-ray burst observed in association with rocket-triggered lightning. *Geophysical Research Letters*, *31*, L05119. https://doi.org/10.1029/2003GL018771

Dwyer, J. R., Smith, D. M., & Cummer, S. (2012). High-energy atmospheric physics: Terrestrial gamma-ray flashes and related phenomena. *Space Science Reviews*, *173*, 133–196. https://doi.org/10.1007/s11214-012-9894-0

Dwyer, J. R., Smith, D. M., Hazelton, B. J., Grefenstette, B. W., Kelley, N. A., Lowell, A. W., et al. (2015). Positron clouds within thunderstorms. *Journal of Plasma Physics*, *81*(475810405). https://doi.org/10.1017/S0022377815000549

Eack, K. B., & Beasley, W. H. (2015). Long-duration X-ray emissions observed in thunderstorms. *Journal of Geophysical Research: Atmospheres*, *120*, 6887–6897. https://doi.org/10.1002/2015JD023262

Eack, K. B., Beasley, W. H., Rust, W. D., Marshall, T. C., & Stolzenburg, M. (1996a). X-ray pulses observed above a mesoscale convective system. *Geophysical Research Letters*, *23*(21), 2915–2918.

Eack, K. B., Beasley, W. H., Rust, W. D., Marshall, T. C., & Stolzenburg, M. (1996b). Initial results from simultaneous observation of X-rays and electric fields in a thunderstorm. *Journal of Geophysical Research*, *101*(D23), 29,637–29,640.

Eack, K. B., Suszcynsky, D. M., Beasley, W. H., Roussel-Dupre, R., & Symbalisty, E. (2000). Gamma-ray emissions observed in a thunderstorm anvil. *Geophysical Research Letters*, *27*(2), 185–188.

Fishman, G. J., Baht, P. N., Mallozzi, R., Horack, J. M., Koshut, T., Kouveliotou, C., et al. (1994). Discovery of intense gamma-ray flashes of atmospheric origin. *Science*, *164*, 1313.

Gurevich, A. V., Milikh, G. M., & Roussel-Dupré, R. (1992). Runaway electron mechanism of air breakdown and preconditioning during a thunderstorm. *Physics Letters A*, *165*(5-6), 463–468. https://doi.org/10.1016/0375-9601(92)90348-P

Hutchins, M. L., Holzworth, R. H., Brundell, J. B., & Rodger, C. J. (2012). Relative detection efficiency of the World Wide Lightning Location Network. *Radio Science*, *47*, RS6005. https://doi.org/10.1029/2012RS005049

Kelley, N. A., Smith, D. M., Dwyer, J. R., Splitt, M., Lazarus, S., Martinez-McKinney, F., et al. (2015). Relativistic electron avalanches as a thunderstorm discharge competing with lightning. *Nature Communication*, *6*, 7845.

Kochkin, P. O., Sarria, D., Skeie, C., van Deursen, A. P. J., deBoer, A. I., Barnet, M., et al. (2018). In-flight observation of positron annihilation by ILDAS. *Journal of Geophysical Research: Atmospheres*, *123*, 8074–8090. https://doi.org/10.1029/2018JD028337

Kochkin, P. O., van Deursen, A. P. J., Marisaldi, M., Ursi, A., deBoer, A. I., Barnet, M., et al. (2017). In-flight observation of gamma ray glows by ILDAS. *Journal of Geophysical Research: Atmospheres*, *122*, 12,801–12,811. https://doi.org/10.1002/2017JD027405

Lehtinen, N. G., Bell, T. F., & Inan, U. S. (1999). Monte Carlo simulation of runaway MeV electron breakdown with application to red sprites and terrestrial gamma flashes. *Journal of Geophysical Research*, *104*(A11), 24,699–24,712.

Mach, D. M. (2015). Technique for reducing the effects of nonlinear terms on electric field measurements of electric field sensor arrays on aircraft platforms. *Journal of Atmospheric and Oceanic Technology*, *32*, 993–11,003. https://doi.org/10.1175/JTECH-D-14-00029.1

Mach, D. M., & Koshak, W. J. (2007). General matrix inversion technique for the calibration of electric field sensor arrays on aircraft platforms. *Journal of Atmospheric and Oceanic Technology*, *24*, 1576–1587. https://doi.org/10.1175/JTECH2080.1

McCarthy, M. P., & Parks, G. K. (1985). Further observations of X-rays inside thunderstorms. *Geophysical Research Letters*, *12*(6), 393–396.

McCarthy, M. P., & Parks, G. K. (1992). On the modulation of X-ray fluxes in thunderstorms. *Journal of Geophysical Research*, *97*(D5), 5857–5864.

Moore, C. B., Each, K. B., Aulich, G. D., & Rison, W. (2001). Energetic radiation associated with lightning stepped-leaders. *Geophysical Research Letters*, *28*(11), 214–2144.

Østgaard, N., Balling, J. E., Bjørnsen, T., Brauer, P., Budtz-Jørgensen, C., Bujwan, W., et al. (2019). The Modular X- and Gamma- ray Sensor (MXGS) of the ASIM payload on the International Space Station. *Space Science Reviews*, *215*, 28. https://doi.org/10.1007/s11214-018-0573-7

Parks, G. K., Mauk, B. H., Spiger, R., & Chin, J. (1981). X-ray enhancements detected during thunderstorms and lightning activity. *Geophysical Research Letters*, *8*(11), 1176–1179.

Picone, J. M., Hedin, A. E., Drob, D. P., & Aikin, A. C. (2002). NRLMSISE-00 empirical model of the atmosphere: Statistical comparisons and scientific issues. *Journal of Geophysical Research*, *107*(A12), 1468. https://doi.org/10.1029/2002JA009430

Rakov, V. A., & Uman, M. A. (2003). *Lightning physics and effects*. New York: Cambridge University Press.

Rison, W., Krehbiel, P. R., Thomas, R. J., Rodeheffer, D., & Fuchs, B. (2012). The Colorado lightning mapping array. AGU Fall Meeting Abstracts, AE23B-0319.

Rodger, C. L., Brundell, J. B., & Dowden, R. L. R. L. (2005). Location accuracy of VLF World-Wide Lightning Location (WWLL) network: Post-algorithm upgrade. *Annales Geophysicae*, *23*, 177–290. https://doi.org/10.5194/angeo-23-277-2005

Rutjes, C., Sarria, D., Broberg Skeltved, A., Luque, A., Diniz, G., Østgaard, N., & Ebert, U. (2016). Evaluation of Monte Carlo tools for high energy atmospheric physics. *Geoscientific Model Development*, *9*, 3961–3974. https://doi.org/10.5194/gmd-9-3961-2016

Rutledge, S., Reimel, K., Fuchs, B., & Xu, W. (2017). GLM validation studies in Colorado (and a brief look at the tropics). AGU Fall Meeting, AE41A-05.

Sarria, D., Rutjes, C., Diniz, G., Luque, A., Ihaddadene, K. M. A., Dwyer, J. R., et al. (2018). Evaluation of Monte Carlo tools for high energy atmospheric physics II: Relativistic runaway electron avalanches. *Geoscientific Model Development Discussions*, *2018*, 1–30. https://doi.org/10.5194/gmd-2018-119

Sato, T., Yasuda, H., Niita, K., Endo, A., & Sihver, L. (2008). Development of PARMA: PHITS-based Analytical Radiation Model in the Atmosphere. *Radiation Research*, *170*, 244–259. https://doi.org/10.1667/RR1094.1

Skeltved, A. B., Østgaard, N., Carlson, B., Gjesteland, T., & Celestin, S. (2014). Modeling the relativistic runaway electron avalanche and the feedback mechanism with GEANT4. *Journal of Geophysical Research: Space Physics*, *119*, 9174–9191. https://doi.org/10.1002/2014JA020504

Smith, D. M., Bowers, G. S., Kamogawa, M., Wang, D., Ushio, T., Ortberg, J., et al. (2018). Characterizing upward lightning with and without a terrestrial gamma ray flash. *Journal of Geophysical Research: Atmospheres*, *123*, 11,321–11,332. https://doi.org/10.1029/2018JD029105

Tsuchiya, H., Enoto, T., Iwata, K., Yamada, S., Yuasa, T., Kitaguchi, T., et al. (2013). Hardening and termination of long-duration $\gamma$ rays detected prior to lightning. *Physical Review Letters*, *111*, 015001. https://doi.org/10.1103/PhysRevLett.111.015001

Tsuchiya, H., Enoto, T., Yamada, S., Yuasa, T., Nakazawa, K., Kitaguchi, T., et al. (2011). Long-duration $\gamma$ rays emissions from 2007 and 2008 winter storms. *Journal of Geophysical Research*, *116*, D09113. https://doi.org/10.1029/2010JD015161

Wilson, C. T. R. (1925). The acceleration of $\beta$-particle in strong electrical fields of thunderclouds. *Proceedings of the Cambridge Philosophical Society*, *22*, 534–538.